\documentstyle[twocolumn,prl,aps]{revtex}

\begin{document}
\draft
\preprint{}
\title{Critical Properties in Dynamical Charge Correlation \\           
       Function for the One-Dimensional Mott Insulator}
\author{ Tatsuya Fujii and Norio Kawakami } 
\address{Department of Applied Physics, Osaka University, Suita, Osaka 565, 
Japan} 
\date{\today}
\maketitle
\begin{abstract}
Critical properties in the dynamical charge correlation 
function for the one-dimensional Mott insulator   
are studied. By properly taking into account {\it the 
final-state interaction} between the charge and spin degrees of freedom, 
we find that the edge singularity in the charge
correlation function is governed by massless 
spinon excitations, although it is naively expected that 
spinons do not directly contribute to the charge excitation
over the Hubbard gap. 
We obtain the momentum-dependent anomalous critical  exponent by 
applying the finite-size scaling analysis to the 
Bethe ansatz solution of the half-filled Hubbard model. 
\end{abstract}
\section{Introduction}
Recently strongly correlated electron systems 
in one dimension (1D) have  attracted much interest. 
In particular, charge and spin excitations for
the Mott insulator have been intensively studied.
Various experimental methods, by which 
the dynamical correlation functions are directly observed, 
have revealed  striking properties 
in the Mott insulator. For example, the recent photoemission
experiments\cite{ex5,ex6,ex7} for the 1D compounds 
SrCuO$_{2}$, Sr$_{2}$CuO$_{3}$ and NaV$_{2}$O$_{5}$ have clarified
dynamical properties of a {\it single hole} doped in the 1D Mott 
insulator. These experiments have indeed stimulated extensive 
theoretical studies on the one-particle Green function for 
1D correlated systems, which have been done numerically 
\cite{shiba,preuss,haas,ppenc,maekawa,loren,mila,zacher,ogata}
and analytically.\cite{S&P1,meden,voit2,S&P2,voit,nagaosa,fujii}
The overall feature of the spin-charge separation found 
in the photoemission experiments
has been explained rather well.\cite{maekawa}
In this connection, it has been claimed that 
{\it the final-state interaction} between the charge and spin 
degrees of freedom plays the crucial role to determine 
the edge singularity in the spectrum, based on the conformal 
field theory (CFT) analysis.\cite{S&P2,fujii}

The dynamical  charge correlation function is
another key quantity to explore the characteristics of the 
Mott insulator, which is particularly important to 
analyze  optical experiments. This quantity has also been  
studied intensively so far for 1D correlated electron systems.
\cite{tsune,Tohyama,Mori,augier,Geb,Carmelo} 
For example, the recent theoretical treatment
of a metallic system close to the Mott transition\cite{Mori} 
has clarified  that the change of the weight
in the dynamical charge correlation function clearly describes the 
characteristics of the metal-insulator transition. 

Motivated by the above investigations, in this paper we study
critical properties in the dynamical charge correlation 
function for the 1D Mott insulator.
For this purpose, we consider the 1D half-filled Hubbard model 
as a Mott insulator and use the exact solution of
the Bethe ansatz method.\cite{LiebWu,taka,woy2}
Applying CFT techniques, we analyze  the exact finite-size spectrum 
to obtain the critical exponents for the  dynamical 
charge correlation function exactly. We clarify how the edge 
singularity in the massive charge excitation spectrum 
over the Hubbard gap is controlled by massless spinon excitations
by properly taking into account  {\it the final-state interaction} 
between the charge and spin degrees of freedom. 
In particular, we point out that a particle-hole charge excitation
is considered to act as {\it two mobile impurities} in massless spinons, 
and the resulting scattering phase shifts are the essential quantities 
to determine the edge singularity.

This paper is organized as follows. In {\S} 2, based on the
exact solution of the 1D Hubbard model,\cite{LiebWu} we   
present the basic formulation following refs. 27 and 28 
to investigate a particle-hole excitation over the Hubbard gap. 
Then in {\S} 3 we calculate the exact finite-size spectrum, 
and derive the edge singularity in  the charge correlation function
by employing the finite-size scaling idea in CFT. 
We discuss the anomalous critical behavior in the singularity  
by evaluating the momentum-dependent critical exponents. 
Brief summary is given in {\S} 4.

\section{Charge Excitations in the Mott Insulator}\label{sec:1}

Let us start with the ordinary 1D repulsive Hubbard model, 
\begin{eqnarray}
{\cal H}
=-t{\sum_{i}}{\sum_{\sigma}}
(c^{\dagger}_{i,\sigma}c_{i+1,\sigma}
                          +c^{\dagger}_{i+1,\sigma}c_{i,\sigma}) 
 +U{\sum_{i}} n_{i\uparrow}n_{i\downarrow}\cr
 -\frac{h}{2} {\sum_{i}}(n_{i\uparrow}-n_{i\downarrow}), \cr
\label{eqn:de-1}
\end{eqnarray}
where $c^{\dagger}_{i,\sigma}$ is the creation operator for electrons  
and $h$ is a magnetic field. We henceforth set $t=1$ for simplicity.
The exact solution of this model\cite{LiebWu} is given by  
a set of the Bethe equations for the charge rapidities 
$k_{j}$ and the spin rapidities $\lambda_{\alpha}$,
\begin{eqnarray}
&&   N k_{j} = 2 \pi {\cal I}_{j}
             -\sum_{\beta=1}^{M}  
             2 {\rm tan}^{-1} [\frac{\sin k_{j}-\lambda_{\beta}}{c}] \cr
%
&& \sum_{j=1}^{N_{e}} 2 {\rm tan}^{-1} [\frac{\lambda_{\alpha}-\sin 
k_{j}}{c}]
    = 2\pi {\cal J}_{\alpha} 
      + \sum_{\beta=1}^{M} 2{\rm tan}^{-1} 
         [\frac{\lambda_{\alpha}-\lambda_{\beta}}{2c}], 
\label{eqn:de-2}
\nonumber
\end{eqnarray}
where we set $c=U/4$ $(>0)$. $N_{e}$ and $M$ represent 
the number of electrons and  down spins, and
$N$ is the number of the lattice sites.
For the repulsive case $U>0$,
the rapidities are real numbers for the ground state 
as well as for excited states within the lower Hubbard band,
which are classified by the set of the quantum numbers 
${\cal I}_{j}$ and ${\cal J}_{\alpha}$ for 
the charge and spin sectors. 

\subsection{Particle-hole excitations at half-filling}

For the half-filled case ($N_e=N$)
the available quantum numbers ${\cal I}_{j}$ for the charge sector 
are completely filled, giving rise to  
the incompressible state of the Mott insulator.
So, charge excitations over the Hubbard gap, which
are necessary  for the evaluation of the dynamical charge 
correlation function, cannot be described by real rapidities, 

It is known that the charge excitations of particle-hole type
are specified by {\it complex} charge rapidities 
in bound pairs.\cite{taka}  Following the methods of 
Woynarovich,\cite{woy2} 
we briefly summarize how to treat such complex rapidities.
As a simple charge excitation at half-filling, we here introduce 
a pair of complex charge rapidities, 
which are denoted as $k^{\pm}=\kappa \pm i \chi$ 
with the real quantities  $\kappa$ and $\chi$. 
In the thermodynamic limit, these charge rapidities are 
coupled with the corresponding spin rapidity $\Lambda$ 
to yield a bound pair,\cite{taka} 
\begin{eqnarray}
  \sin(\kappa \pm {\rm i} \chi)=\Lambda \mp {\rm i} c.
\end{eqnarray}
Since this pair adds two particles to the system. 
we need to introduce two holes, labeled
by the real charge rapidities $k_{l}$ and $k_{m}$,
in order to  keep the particle number unchanged. 
We thus obtain the Bethe equations for a particle-hole
excitation over the Hubbard gap,\cite{woy2}
\begin{eqnarray}
     N k_{j} {\hspace{-1mm}}= {\hspace{-1mm}} 2 \pi {\cal I}_{j} 
       {\hspace{-1mm}} - {\hspace{-1.5mm}}
     \sum_{\beta=1}^{M-1}  
      2 {\rm tan}^{-1} [\frac{\sin k_{j} - \lambda_{\beta}}{c}] 
     {\hspace{-1mm}} - {\hspace{-1mm}} 
     2 {\rm tan}^{-1} [\frac{\sin k_{j} - \Lambda}{c}]    \cr
     \sum_{j \neq l,m} 2 {\rm tan}^{-1} 
               [\frac{\lambda_{\alpha}-\sin k_{j}}{c}]
      {\hspace{-1mm}}={\hspace{-1mm}}
      2\pi {\cal J}_{\alpha}^{'} 
      {\hspace{-1mm}}+{\hspace{-1mm}}
      \sum_{\beta=1}^{M-1}2{\rm tan}^{-1}
      [\frac{\lambda_{\alpha}-\lambda_{\beta}}{2c}]. \cr
&
\label{eqn:de-3}
\end{eqnarray}
These equations are subject to the constraints
for the momentum conservation, 
\begin{eqnarray}
  N {\kappa} = 2 \pi {\cal I} -
 \frac{\pi}{2}\sum_{\beta=1}^{M-1} {\rm sign} (\Lambda-\lambda_{\beta}) 
-\sum_{\beta=1}^{M-1} {\rm tan}^{-1} [\frac{\Lambda - \lambda_{\beta}}{2c}] 
\cr
\sum_{j \neq l,m} 2{\rm tan}^{-1} [\frac{\Lambda-\sin k_{j}}{c}]
        = 2\pi {\cal J} + \sum_{\beta=1}^{M-1} 2{\rm tan}^{-1} 
                             [\frac{\Lambda-\lambda_{\beta}}{2c}],
        \cr
\label{eqn:de-4}
\end{eqnarray}
which are recast into more convenient form,
\begin{eqnarray}
  \Lambda=\frac{1}{2}(\sin k_{l}+\sin k_{m}).
\label{eqn:de-5}
\end{eqnarray}
Note that the unknown parameters for excitations are 
now reduced to two rapidities $k_{l}$ and $k_{m}$ 
thanks to the above constraints, which naturally  reproduces
the fact that a particle-hole excitation 
is classified by two independent momentums.
Eqs. (\ref{eqn:de-3}) and (\ref{eqn:de-5})
are our starting equations to analyze the 
finite-size corrections to a particle-hole excitation 
over the Hubbard gap and to get further information on the charge 
correlation function. 

\subsection{Dispersion relations in magnetic fields}

Before computing the finite-size corrections,
we first summarize the results on the dispersion relation
by slightly extending the method of ref. 28 
to include the effect of magnetic fields.
Following a standard way, let us introduce the distribution 
functions for the charge and spin sectors, 
which are denoted as $\rho (k)$ and $\sigma (\lambda)$, respectively.
\begin{eqnarray}
  \rho(k)=\frac{1}{2\pi}+\int_{-\lambda_{0}}^{\lambda_{0}} 
          \frac{{\rm d}\lambda}{2\pi}
          \cos k K_{12}(\sin k-\lambda) \sigma(\lambda) {\hspace{1.5cm}} \cr
         +\frac{1}{2\pi N}\cos k K_{12}(\sin k-\Lambda)  \cr
  \int_{-\pi}^{\pi} \frac{{\rm d}k}{2\pi} K_{21}(\lambda-\sin k) \rho(k)
  - \frac{1}{2\pi N} \sum_{\alpha =l,m}K_{21}(\lambda-\sin k_{\alpha}) \cr
  = \sigma(\lambda)
  + \int_{-\lambda_{0}}^{\lambda_{0}}
   \frac{{\rm d}\lambda^{'}}{2\pi} K_{22}(\lambda-\lambda^{'}) 
\sigma(\lambda^{'})
   \cr
\label{eqn:de-5.5}
\end{eqnarray}
where
$K_{12}(x)$$=$$K_{21}(x)=2c/(c^{2}+x^{2})$,
$K_{22}(x)$$=$$4c/(4c^{2}+x^{2})$. 
Note that a particle-hole excitation is incorporated in
the integral equations 
 through the terms of order of $1/N$ including 
$k_l$  and $k_m$. 

The energy and momentum for this excited state are given by, 
\begin{eqnarray}
E &=& \sum_{j {\neq} l,m}(h_{c}-2\cos k_{j})+\sum_{\alpha=1}^{M}h_{s} \cr
    & & +(h_{c}-2\cos k^{+}+h_{c}-2\cos k^{-}), \cr
    & & \cr
P &=& \sum_{j\neq l,m}k_{j} + (k^{+}+k^{-}),
\label{eqn:de-6}
\end{eqnarray}
where $h_{c}=-h/2$ and $h_{s}=h$. 
Using (\ref{eqn:de-5.5}), we recast them to
simple formulas,  
\begin{eqnarray}
\omega_c \equiv E - E_{0}&=&  
U-\varepsilon_{c}(k_{l})-\varepsilon_{c}(k_{m}) ,\cr
          & & \cr
q_c \equiv P - P_{0}&=& -p(k_{l})-p(k_{m}),
\label{eqn:su-7}
\end{eqnarray}
where $E_{0}$ and $P_{0}$ is the energy and momentum of 
the ground state. Here we have introduced the dressed energy 
$\varepsilon_{c}(k)$ and the dressed momentum $p(k)$
for the simplest charge excitation (often referred to as holon),
which are given by
\begin{eqnarray}
  \varepsilon_{c}(k)&=&-2\cos k-2\int_{0}^{\infty}
            \frac{{\rm e}^{-c\omega}}{\omega \cosh c\omega}
            J_{1}(\omega)\cos(\omega \sin k) {\rm d}\omega \cr
           &&{\hspace{0.8cm}} -\int_{|\lambda |>\lambda_{0}} 
                \frac{1}{4c} \cosh^{-1} \frac{\pi}{2c(\sin k-\lambda)} 
                     \varepsilon _{s}(\lambda)
                           {\rm d}\lambda ,\cr
  p(k)&=& k+\int_{0}^{\infty} \frac{{\rm e}^{-c\omega}}
                                             {\omega \cosh c\omega}
                  J_{0}(\omega) \sin (\omega \sin k){\rm d}\omega \cr
      &&{\hspace{3cm}}- \int_{|\lambda |>\lambda_{0}} 
                         G(\sin k-\lambda) \sigma(\lambda), \cr
&&
\label{eqn:de-8}
\end{eqnarray}
where $J_{n}(\omega)$ is the $n$-th order Bessel function. 
This charge excitation is coupled with spin excitation (spinon), 
whose energy $\varepsilon _{s}$ is 
given by the solution to the following integral equation,
\begin{eqnarray}
   \varepsilon _{s}(\lambda)= \varepsilon _{s}^{0}(\lambda)-
                         \int_{-\lambda_{0}}^{\lambda_{0}} 
                         \frac{{\rm d}\lambda ^{'}}{2\pi}
                            K_{22}(\lambda-\lambda^{'}) 
                          \varepsilon _{s}(\lambda^{'})
                              \cr
\label{eqn:de-9.1}
\end{eqnarray}
where,
\begin{eqnarray}
   \varepsilon _{s}^{0}(\lambda)&=&h-2\int^{\pi}_{-\pi} 
                          \frac{{\rm d} k}{2\pi}
                              \cos ^{2}k K_{12}(\sin k-\lambda) ,\cr
   G(\lambda)&=& \int_{0}^{\infty} \frac{\sin \omega \lambda}
                               {\omega \cosh c\omega} {\rm d} \omega.
\label{eqn:de-9.0}
\end{eqnarray}
The dispersion for the particle-hole excitation spectrum is described 
by $\omega_c(q)$, which includes the effects of  
magnetic fields.

In Fig. 1 we show the charge excitation spectrum ($q=q_c$)
for a given $U$ in several choices of magnetic fields.
The charge excitation spectrum distributes in the continuum over
the Hubbard gap. It is seen from the figure that
the lower edge of the spectrum changes its character 
at the critical momentum $q=\pi \pm q_L(h)$. For 
$\pi -q_L<q<\pi +q_L$,
the lower edge is given by changing a rapidity 
$k_{m}$ with the other being fixed as $k_{l}=-\pi$.
This excitation corresponds to a particle-hole
excitation  from the top of the lower Hubbard band to
an excited state in the upper Hubbard  band.  On the other hand, 
for 
$0<q<\pi -q_L$ and $\pi +q_L<q<2\pi$, the lower edge is featured by the 
excitation
with $k_l=k_m$, which makes a particle (hole) with the momentum $q/2$
($-q/2$) in the the upper (lower) Hubbard band.
So, the corresponding critical behavior of the edge
singularity is different between these two regimes, which 
will be explicitly studied in the next section.

\begin{figure}[h]
\caption{
Charge excitation spectrum for $U=6$ in several choices of 
magnetic fields: (a), (b) and (c) correspond to the case 
of the magnetization $m=0,0.5 {\hspace{1mm}}{\rm and} {\hspace{1mm}} 1$. 
The excitations are allowed for the hatched region. 
The lower edge of spectrum changes its character at the critical momentum 
$\pi \pm q_L$. In the case of $U \rightarrow \infty$ or $m \rightarrow 1$, 
$q_L$ equals to $0$ in $0<q<2\pi$. On the other hand, $q_L$ becomes $\pi$ 
at $U=0$. 
}
\label{fig:1}
\end{figure}
%
%
\section{Critical Behavior in Dynamical Charge Correlation Function}

We now consider the dynamical charge correlation function 
for the  Mott insulator and show  that the anomalous power-law 
behavior appears in the spectrum,  which is controlled by massless
spinon excitations subject to the two phase shifts due
to a particle-hole excitation.

Let us start with the  charge correlation function, 
\begin{eqnarray}
 D(x,t)&=&<n(x,t),n(0,0)> \cr
&& \cr
 &=&\sum_{\mu} |<0|n(0)|\mu>|^{2} 
                               {\rm e}^{{\rm i}\Delta P_{\mu}x
                                     -{\rm i}\Delta E_{\mu}t} \cr
                &\simeq& \sum_{\{k_{l},k_{m}\}} 
              {\rm e}^{{\rm i} q_cx-{\rm i}
                 \omega_ct} D_{s}(x,t),
\label{eqn:de-14}
\end{eqnarray}
where $n=\sum_{\alpha}n_{\alpha}$ is the charge density operator. 
In the third line of the equation, we have approximately substituted  
$\omega_c(k_l,k_m)$ and $q_c(k_l,k_m)$ given by (\ref{eqn:su-7})
for the energy and the momentum of excited states. 
Since the action of the charge density operator 
$n$ creates not only particle-hole charge excitations, but
also induces low-energy spin excitations, the contribution
from the spin sector to $<n(x,t)n(0,0)>$ is denoted by 
the spin correlator $D_{s}(x,t)$.  In the following 
we will mainly deal with  the Fourier transform of 
the correlation function,
\begin{eqnarray}
&  D(q,\omega)\simeq  \sum_{\{k_{l},k_{m}\}} 
     \int {\rm d}x \int {\rm d}t {\hspace{1mm}}
                    {\rm e}^{{\rm -i} (q-q_c)x +{\rm i}
           (\omega-\omega_c)t} D_{s}(x,t), \cr
&
\label{eqn:de-11}
\end{eqnarray}
for which the main spectrum is featured by  
the massive charge excitation $\omega_c(q)$ while its critical 
behavior is essentially determined by the spin correlator $D_{s}(x,t)$.
It is now clear that the long-time or large-distance 
behavior of  $D_{s}(x,t)$ is important to discuss 
the critical behavior of the dynamical charge correlation function.
Therefore our remaining task is to study 
low-energy properties of $D_{s}(x,t)$ exactly when a particle-hole
excitation is created.

\subsection{Finite-size spectrum and conformal properties}

In order to apply the methods developed in CFT\cite{bpz} to
the evaluation of $D_s(x,t)$, 
let us first compute the finite-size corrections to 
the excitation energy.\cite{cardy,affleck}
We recall here that in the calculation of the spectrum
in the previous section, 
we have not taken into account the fact that massless spinons are
scattered by two dressed particles created 
for the charge part (holons), and have discarded 
the resulting scattering phase shifts in the 
thermodynamic limit.  However, it turns  out  
that this coupling between the charge and spin degrees of freedom 
is essential to determine anomalous critical properties in the charge 
correlation function.  Since we are now considering the 
situation that a particle-hole pair excitation is suddenly created,
this interaction may be regarded as the so called 
{\it final-state interaction}. We show explicitly the above fact 
by correctly evaluating the final-state interaction.

We can perform the calculations of the finite-size spectrum 
by extending those previously  done for
the photoemission spectrum.\cite{fujii} 
In order to deal with the finite-size effects, 
let us  rewrite the coupled integral equations of 
distribution functions (\ref{eqn:de-5.5}),
\begin{eqnarray}
  \rho(k)=\frac{1}{2\pi}+\int_{\lambda^{-}}^{\lambda^{+}} 
        \frac{{\rm d}\lambda}{2\pi}
        \cos k K_{12}(\sin k-\lambda) \sigma(\lambda) {\hspace{1.5cm}} \cr
       +\frac{1}{2\pi N}\cos k K_{12}(\sin k-\Lambda) ,\cr
  \int_{-\pi}^{\pi} \frac{{\rm d}k}{2\pi} K_{21}(\lambda-\sin k) \rho(k)
  -\frac{1}{2\pi N} \sum_{\alpha =l,m}K_{21}(\lambda-\sin k_{\alpha}) \cr
  = \sigma(\lambda)
  +\int_{\lambda^{-}}^{\lambda^{+}}
   \frac{{\rm d}\lambda^{'}}{2\pi} K_{22}
(\lambda-\lambda^{'}) \sigma(\lambda^{'}), \cr
&
\label{eqn:de-12.3}
\end{eqnarray}
where we have introduced the asymmetric cutoffs for
spin rapidities,  $\lambda^{\pm}$, which are essential to
include the final-state interaction.
Namely, two massive holons created give rise to the 
phase shifts in massless spinons, 
which naturally induce the change in cutoffs
in the order of $1/N$ (compare them with (\ref{eqn:de-5.5})). 
The importance of this effect has been previously noticed  for the 
photoemission spectra in the Mott insulator.\cite{S&P2,fujii} 
In this way, although it is naively expected that 
the massless spinons do not directly coupled with 
the charge excitation, the edge singularity in the charge 
correlation function is affected 
by spinons via the final-state interaction. 

Let us now exactly analyze the excitation energy, 
including the contribution 
from low-energy massless spinon excitations. 
By applying  the computational techniques presented in
\cite{fujii} to eqs.(\ref{eqn:de-12.3}), we 
straightforwardly end up with the following formulas, 
\begin{eqnarray}
   \Delta E &=& \Delta \epsilon _{c}+\Delta \epsilon _{s} \cr
            &=&  U -\varepsilon_{c}(k_{l})-\varepsilon_{c}(k_{m}) 
                 +\frac{2\pi v_{s}}{N}(x+N_{+}+N_{-}), \cr
&&
\label{eqn:de-13}
\end{eqnarray}
where $v_{s}$ is the velocity of massless spinons, 
\begin{eqnarray}
   v_{s}=\frac{\varepsilon^{'}_{s}(\lambda)}
                {2\pi \sigma (\lambda)} 
                  \left |_{\lambda=\lambda_{0}} \right. . 
\label{eqn:de-13.1}
\end{eqnarray}
Note that the first term of the order of unity in 
(\ref{eqn:de-13}) represents the energy $\omega_c$ for
the massive charge excitation, which is 
referred to as the surface energy in boundary CFT.
Now, according to the finite-size scaling idea in CFT,
\cite{cardy,affleck} 
the scaling dimension $x$ for the spin sector is read from the $1/N$ 
corrections 
to the excitation energy, which is obtained as,
\begin{eqnarray}
   x=\frac{1}{4 \xi_{s}^{2}} (-1+n_{c}(k_{l})+n_{c}(k_{m}))^{2} +
             \xi_{s}^{2} (d_{c}(k_{l})+d_{c}(k_{m}))^{2}. \cr
\label{eqn:de-13.2}
\end{eqnarray}
This formula for the scaling dimension is typical for Tomonaga-Luttinger 
liquids classified by $c=1$ CFT.  
Here, the quantity $\xi_{s}=\xi_{s}(\lambda_{0})$ (referred to as 
the dressed charge),  
\begin{eqnarray}
   \xi_{s}(\lambda)= 1- \int_{-\lambda_{0}}^{\lambda_{0}} 
                               \frac{{\rm d}\lambda ^{'}}{2\pi}
                    K_{22}(\lambda-\lambda^{'}) \xi_{s}(\lambda ^{'}),
\label{eqn:de-13.3}
\end{eqnarray}
features the U(1) critical line of $c=1$ CFT 
when we change the strength of interaction or the magnetic field. 
A remarkable point in (\ref{eqn:de-13.2}) is that
there exist two kinds 
of phase shifts $n_{c}$ and $d_{c}$, which 
are caused by the final-state interaction
between the charge and spin degrees of freedom. 
These phase shifts are explicitly obtained as
\begin{eqnarray}
  n_{c}(k_{\alpha}) &=& \int_{-\lambda_{0}}^{\lambda_{0}}
                      \sigma_{\alpha} (\lambda) , \cr
   d_{c}(k_{\alpha})&=&- \frac{1}{2}
                \left(
                 \int_{\lambda_{0}}^{\infty} 
                 \sigma _{\alpha} (\lambda) 
                -
                 \int^{-\lambda_{0}}_{-\infty} 
                 \sigma _{\alpha} (\lambda) 
                \right),
\label{eqn:de-13.4}
\end{eqnarray}
where
\begin{eqnarray}
\sigma _{\alpha}(\lambda){\hspace{-1mm}}= {\hspace{-1mm}}
                          \frac{1}{2\pi}K_{21}(\lambda -\sin k_{\alpha})
                          {\hspace{-1mm}}- {\hspace{-1mm}}
                           \int_{-\lambda_{0}}^{\lambda_{0}} 
                             \frac{{\rm d}\lambda ^{'}}{2\pi}
                               K_{22}(\lambda-\lambda^{'}) 
                                \sigma _{\alpha}(\lambda ^{'}). \cr
\label{eqn:de-13.5}
\end{eqnarray}
It is thus seen that although the scaling dimension $x$ for massless 
spinons is typical for $c=1$ CFT,\cite{fk,kawakami} the massive charge 
sector also contributes to $x$ via the phase shifts $n_{c}$ and $d_{c}$.
In this sense, (\ref{eqn:de-13.2}) is classified as 
{\it shifted} $c=1$ CFT, whose fixed point is different 
even from that of the static impurity problem, as pointed out by 
Sorella and Parola.\cite{S&P2} 
Since in the present case, these phase shifts are considered to be 
caused by the {\it two holons} created, the fixed point belongs to 
Tomonaga-Luttinger liquid with {\it two mobile 
impurities}.\cite{tsukamoto} 
It will be shown that these phase shifts are the key quantities 
to control the anomalous low-energy properties in the 
correlation function. 

Exploiting finite-size scaling techniques in CFT, 
we can now  write down the asymptotic form for
the spin correlation function as 
\begin{eqnarray}
  D_{s}(x,t) \simeq \frac{1}
         {(x-v_{s}t)^{2\Delta_{s}^{+}}(x+v_{s}t)^{2\Delta_{s}^{-}}},
\label{eqn:de-15}
\end{eqnarray}
where $\Delta_{s}^{\pm}$ are conformal dimensions 
for the spin sector, which are related to the scaling dimension 
as $x=\Delta_{s}^{+}+\Delta_{s}^{-}$. 
Thus we can read the low-energy behavior in $D_s(x,t)$ 
correctly from the scaling dimension (\ref{eqn:de-13.2}).

\subsection{Anomalous critical behavior}

Let us now discuss the critical properties
in the dynamical correlation function
by substituting the above results for $D_s(x,t)$ to (\ref{eqn:de-11}). 
As noted in the previous section, 
there appear two different lower edges in the charge spectrum
depending on the momentum $q$, so 
we deal with two cases separately. 
\par

\subsubsection{Edge singularity for $\pi -q_L<q<\pi +q_L$}
We now discuss the critical properties 
along the lower edge of particle-hole excitation spectrum.
Let us start with the case  for $\pi -q_L<q<\pi +q_L$ shown in Fig. 2.
As shown in the previous section, the lower edge is given by
$\omega_c(q)=U-\varepsilon(-\pi) - \varepsilon(k_m)$.
%
In this case, if we neglect the contribution from 
the spin sector, there does not exist the power-law
edge singularity in the charge correlation function, 
so that the anomalous behavior 
around the edge may be completely determined by  
spinon excitations. So the Fourier transform of the 
charge correlation function is given by 
\begin{eqnarray}
D(q,\omega) \propto 
   \theta (\omega-\omega_{c}(q))& ( \omega-\omega_{c}(q))^{X_1(q)}, 
\label{eqn:de-16}
\end{eqnarray}
for $\omega \simeq \omega_{c}(q)$.
The corresponding critical exponent $X_1(q)$ 
directly follows from the scaling dimension 
for the spin sector as
\begin{eqnarray}
X_1(q)=2x-1, 
\end{eqnarray}
where $x$ is given by (\ref{eqn:de-13.2}).
We emphasize here that the critical exponent is {\it dependent on the 
momentum $q$}. Recall that this anomalous behavior for the critical 
exponent is due to the two phase shifts in $x$, which is caused by
the final-state interaction.  In Fig. 2,
we show the results for $X_1(q)$ as a function of $q$ 
for a given  magnetic field.
In the limit of vanishing magnetic fields,
the exponent gradually approaches the value of $-1$ because
the scaling dimension $x$ tends to zero.
On the other hand, exactly  at $h=0$, the power-law singularity
is expected to disappear, which implies that $X_1=0$.
We can show that there indeed exists an energy scale 
which characterizes the   crossover
between these two behaviors. Namely, only in the small region 
close to the edge, $(\omega-\omega_{c}(q))^{(2x-1)} < 1/\Gamma (2x)$,
we can observe a power-law singularity with the exponent $X_1=2x-1$
where $\Gamma$ is the gamma function. When the magnetic field 
is decreased, this region is gradually shrunk because 
$\Gamma (2x) \rightarrow \infty$, and is eventually replaced by 
$D(q,\omega\rightarrow \omega_c) \simeq$constant.
  
\subsubsection{Edge singularity for $0<q<\pi -q_L$ and $\pi +q_L<q<2\pi$}

Let us now turn to the momentum region 
$0<q<\pi -q_L$ and $\pi +q_L<q<2\pi$ in Fig. 1,
in which the lower edge is given by the excitation
with $k_l=k_m$.  Along this edge there occurs a specific 
excitation where a particle (a hole) is created
at symmetric points in the upper (lower) Hubbard band with
the momentum $q/2$ ($-q/2$) for a given $q$.
So, we should be a little bit careful to deduce 
its critical behavior. We start by discarding 
the spinon contributions to clearly see the origin of
the singularity. Let us first recall that for a given $q$
the charge excitation energy $\omega_c$
in (\ref{eqn:su-7}) takes its minimum at 
$k_l=k_m$ when we change the values of $k_l$ and $k_m$.
So, around the edge, $\omega_c$ has the following properties  
for a fixed momentum $q$, 
\begin{eqnarray}
\omega_c^{'}|_{k_{m}=k_{l}} =0, \hspace{3mm} 
\omega_c^{''}|_{k_{m}=k_{l}} > 0 , \cr
\label{eqn:de-12.2}
\end{eqnarray}
which naturally leads to the 
well known square-root behavior like the van Hove singularity,\cite{Mori}
$(\omega-\omega_{c}(q))^{-1/2}$.
This singularity depends on neither the coulomb interaction $U$ 
nor the magnetic field $h$, so long as the spinon 
contribution is neglected.

Now let us take into account the final-state interaction 
between the charge and spin degrees of freedom.
Then low-energy spinons subject to the phase shifts show 
up in the infrared singularity. 
 By properly incorporating
this contribution, we finally end up with the edge singularity
for $0<q<\pi -q_L$ and $\pi +q_L<q<2\pi$,
\begin{eqnarray}
D(q,\omega) \propto 
   \theta (\omega-\omega_{c}(q))& ( \omega-\omega_{c}(q))^{X_2(q)},
\label{eqn:de-17}
\end{eqnarray}
where the corresponding critical exponent is 
\begin{eqnarray}
X_2(q)=-\frac{1}{2}+ \frac{1}{2}(2x-1). 
\end{eqnarray}
The first term of $-1/2$ comes from the square-root
singularity in the dispersion relation, while the latter
contribution reflects low-energy spinon excitations.
In Fig. 2 we show the momentum dependent 
critical exponent for 
$0<q<\pi -q_L$ and $\pi +q_L<q<2\pi$ for several choices 
of magnetic fields.

We note here again that the above exponent seems to approach $-1$
for  vanishing magnetic fields, while the correct exponent 
should be given by  $-1/2$ at $h=0$, corresponding 
to the square-root singularity.  As mentioned above,
there is the crossover between these two behaviors.
In this case, in the  
region  $(\omega-\omega_{c}(q))^{(2x-1)/2} < 1/\Gamma (2x)$,
we can see a power-law behavior with the exponent $X_2=-1/2+(2x-1)/2$,
while for  higher energies, we can see the square root singularity.
 When $h \rightarrow 0$, this region with the exponent $X_2$
is gradually shrunk, being replaced by the square-root 
behavior.

\begin{figure}[h]
\caption{
Momentum-dependent critical exponent $X_{1}(q)$ and $X_{2}(q)$, 
which correspond to   
the lower edge of spectrum in $\pi -q_L<q<\pi +q_L$ and 
$0<q<\pi -q_L,\hspace{1mm} \pi +q_L <q<2\pi$. 
The magnetization is given by 
$m=0,0.5 {\hspace{1mm}}{\rm and}{\hspace{1mm}} 1$ 
with the fixed 
$U=6.$}
\label{fig:2}
\end{figure}
\subsection{Comparison with photoemission spectrum}

As shown above, the edge singularity in the 
dynamical charge correlation function for the 1D
Mott insulator is controlled by massless spinons
subject to the phase shifts caused by the final-state
interaction.
 We recall again that similar anomalous properties have been already 
reported in the study on the photoemission spectrum, for which 
one electron is removed from the system and 
the induced massless spinon excitations play a crucial role for the edge 
singularity.\cite{S&P2,voit,nagaosa,fujii} 
Therefore, the origin of the
anomalous critical behavior in both cases are essentially
same in the sense that their class belongs to the 
Tomonaga-Luttinger liquid with {\it mobile impurities}\cite{tsukamoto}.
It should be  noticed, however,  that in the present case 
for the charge correlation function
two mobile impurities are created as a
particle-hole excitation, while for the photoemission spectra,
only a single mobile impurity is created.
So, the momentum-dependence of the critical exponents are 
quite different between these two cases.

\section{Summary}
We have studied the critical behavior in the dynamical charge 
correlation function for the 1D Mott insulator. 
Anomalous critical exponents have been calculated for the edge 
singularity in the charge excitation spectrum for the half-filled 
Hubbard model, by combining the Bethe ansatz with finite-size scaling 
methods in CFT. It has been claimed that the final-state interaction 
between the charge and spin degrees of freedom plays 
an important role to produce the anomalous critical behavior 
dependent on the momentum. We have further pointed out
that the universality class of the present system belongs to the 
Tomonaga-Luttinger liquid with {\it two mobile impurities}.

In this paper, we have studied the charge density correlation function 
for the  half-filled Hubbard model. Similar analysis 
can be straightforwardly applied  to 
the dynamical {\it spin} correlation function for
spin gapped systems in a metallic phase.  In such cases,
the roles played by the charge and spin degrees of freedom
are interchanged; namely massless holons subject to
the spinon phase shifts essentially determine the 
anomalous critical properties in the spin excitation spectrum.

\section*{Acknowledgements}
We thank Y. Tsukamoto for valuable discussions.
This work was partly supported by a Grant-in-Aid from the Ministry
of Education, Science, Sport and Culture, Japan.
%

%
%
\end{document}